\documentstyle[twocolumn,epsf]{jpsj}

\def\runtitle{Intrinsic Josephson effect}
\def\runauthor {Noriaki {\sc Kawai}, 
Hiroki {\sc Tsuchiura}, Yukio {\sc Tanaka},
 and Jun-ichiro {\sc Inoue}
}

\title
{Intrinsic Josephson Effect in the Layered Two-Dimensional $t$-$J$ Model}  

\author
{
Noriaki {\sc Kawai}$^{1}$,
Hiroki {\sc Tsuchiura}$^{2}$, Yukio {\sc Tanaka}$^{1}$
and Jun-ichiro {\sc Inoue}$^{1}$
}

\inst
{
$^{1}$Department of Applied Physics, Nagoya University, Nagoya, 464-8603,Japan\\
$^{2}$CREST, Japan Science and Technology Corporation (JST), Nagoya, 464-8603, Japan
}

\recdate
{
\today
}

\abst
{
The intrinsic Josephson effect in the high-$T_{c}$ superconductors is
studied using the layered two-dimensional $t$-$J$ model.
The d.c. Josephson current which flows perpendicular to the $t$-$J$ planes
is obtained within the mean-field approximation and the Gutzwiller
approximation.
We find that the Josephson current has its maximum near the optimum doping
region as a function of the doping rate.
}

\kword
{
Two-dimensional $t$-$J$ model, Gutzwiller approximation, 
layered system,
intrinsic Josephson effect
}

\newcommand{\ktd}{k_{\rm 2D}}
\newcommand{\idel}{{\it{\Delta}}}
\begin{document}
\sloppy
\maketitle


%
Since the discovery of high-$T_{\rm c}$ superconductors,
there have been revealed many novel superconducting properties 
which can not be expected for conventional superconductors. 
One of these is the $d$-wave symmetry of the pair potential 
giving rise to 
novel properties of phase coherence which has been confirmed 
by both 
experimental and theoretical studies 
in tunneling and Josephson effects
\cite{Sigrist1,Wollman,Hu,Kashiwaya2,Tanaka1,
Iguchi1,Tanaka94,Sigrist2,Tanaka3,Barash2}. 
The strong two dimensionality and the anomalous transport 
along $c$-axis direction have brought about another  
important feature of the existence of the intrinsic Josephson 
effects in Bi$_{2}$Sr$_{2}$CaCu$_{2}$O$_{8}$(BSCCO) 
\cite{Kleiner1,Kleiner2}. 
It was reported that a small BSCCO single crystal behaved like a 
series of Josephson junctions. 
It is natural to consider a BSCCO superconductor as a stack 
of superconducting sheets consisting of CuO bilayers separated by 
BiO and SrO layers acting as insulators. 
After the discovery of intrinsic Josephson effect by Kleiner 
\cite{Kleiner1,Kleiner2}, 
there are several evidences which support the existence of 
Josephson coupling between adjacent CuO bilayers 
\cite{Itoh,Matsuda1,Matsuda2,Suzuki,Latyshev}. 
Up to now there are several theoretical works about this problem 
from the view point of the layered structure \cite{Tachiki}. 
However, existing theories \cite{Klemm,Radtke,Chen} 
do not treat the two important facts, i.e., 
short coherence length and the strong correlation 
on the equal footing. 
Taking account of the short coherence length effect, 
it is suitable to express BSCCO by the lattice model. 
One of the authors  calculated the intrinsic Josephson current 
in layered attractive Hubbard model within mean field approximations
where the $s$-wave conventional paring is considered\cite{Tanaka}.
Later, Schmitt $et$ $al.$ \cite{Hanke}
calculated the Josephson current 
in the same model 
based on the Quantum Monte Carlo 
Calculation and established the existence of 
intrinsic Josephson current from the numerical calculations 
with much more accuracy. 
However, in these two papers, strong correlation effect 
and the resulting $d$-wave symmetry were not considered. 
To overcome this problem, 
in this paper,
we study the intrinsic Josephson current using 
the layered two-dimensional (2D) $t$-$J$ model,
which includes the important features above mentioned. 
\par
The $t$-$J$ model \cite{Zhang1}
is one of the promising models which explain the
low-energy excitations in high-$T_{\rm c}$ superconductors.
Although the 
analytic solution of this model have not yet been obtained,
phase diagrams as a function of doping rate $\delta$ and a
superexchange interaction $J$ are numerically studied
at $T=0$
for one-dimension \cite{Ogata1} and two-dimension
\cite{Putikka,Dagotto1,Dagotto2,Yokoyama}.
Especially, in the 2D $t$-$J$ model,
the obtained phase diagram
as a function of doping is consistent with actual
high-$T_{\rm c}$ superconductors \cite{H.F.}.
\par
In this letter, we study the intrinsic Josephson effect in the layered
two-dimensional $t$-$J$ model.
The intrinsic Josephson current which flows perpendicular to the $t$-$J$
planes is calculated and the current-phase relations are obtained for
several doping rates. 
The Josephson current 
obtained 
as a function of the doping concentration $\delta$ 
has a  maximum near the optimum doping region. 
\par

The Hamiltonian of the $m$-th $t$-$J$ plane is written as
\begin{eqnarray}
{\cal H}_{t-J}^{m} &=& -t\sum_{\langle i,j \rangle \sigma}
( c^{m\dag}_{i\sigma} c^{m}_{j\sigma} + {\rm h.c.} )
\nonumber \\
& & + \sum_{\langle i,j\rangle }
(J\mib{S^{m}_{i}}\cdot \mib{S^{m}_{j}}
 - \frac{J_{N}}{4}n^{m}_{i}n^{m}_{j} )
\label{tj}
\end{eqnarray}
where the Hilbert space is defined on the subspace without double
occupancy, $i, j$ refer to planar sites on a square lattice, 
and $c_{i\sigma}^{m\dag}, c_{i\sigma}^{m}$ represent
fermion operators within the $m$-th layer.
The spin and the number operators in this layer are defined as
\[
\mib{S^{m}_{i}} = \sum_{\alpha,\beta}c_{i\alpha}^{m\dagger} 
(\frac{1}{2}
\mib{\sigma} )_{\alpha\beta}c^{m}_{i\beta}  ~,
~~n^{m}_{i} = \sum_{\sigma}c^{m\dag}_{i\sigma} c^{m}_{i\sigma}, 
\]
respectively. 
There seems to be an agreement in the conclusion that the $c$-axis conductivity
in the normal state generally cannot be explained as a coherent interlayer
transport.
We assume, however, that the coherency among
the layers is restored in the superconducting state. 
Thus we allow for the electron hopping between the $m$-th and the $m+1$-th 
layers,
\begin{equation}
{\cal H}_{z}^{m} = 
-t_{z}( c^{m+1\dag}_{i\sigma} c^{m}_{i\sigma} + {\rm h.c.} ).
\end{equation}
Now the total Hamiltonian of the layered $t$-$J$ model is written as
\begin{equation}
{\cal H} = \sum_{m}({\cal H}_{t-J}^{m} + {\cal H}_{z}^{m})
-\mu\sum_{m,i,\sigma} c_{i\sigma}^{m\dag}c_{i\sigma}^{m}
\end{equation}
where $\mu$ is the chemical potential of the whole system.

For this Hamiltonian, we consider Gutzwiller-type variational wave functions
$P_{G}|\Phi\rangle$ with 
$P_{G} = \Pi_{m,i}(1 - n_{i\uparrow}^{m}n_{i\downarrow}^{m})$
being the Gutzwiller projection operator which excludes the double occupancy,
and $|\Phi\rangle$ being a one-body mean-field wave function.
Due to the Gutzwiller projection,
it is usually difficult to calculate the variational energy.
Thus, we use a Gutzwiller approximation in which the effect of the projection
is represented by statistical weight factors.
For the two-dimensional $t$-$J$ model, it is known that
this approximation and the variational Monte Carlo simulation give very
similar variational energies \cite{Yokoyama}.

In the Gutzwiller approximation,
the expectation values of the terms in ${\cal H}_{t-J}^{m}$ are estimated as
\begin{eqnarray}
\langle c^{m\dag}_{i\sigma}c_{j\sigma}^{m} \rangle
&=& g_{t}\langle c^{m\dag}_{i\sigma}c_{j\sigma}^{m}\rangle_{0},
\nonumber \\
\langle  \mbox{\boldmath $S_{i}^{m}\cdot S_{j}^{m}$} \rangle
&=& g_{s}\langle \mbox{\boldmath $S_{i}^{m}\cdot S_{j}^{m}$}\rangle_{0},
\nonumber \\
\langle n_{i}^{m}n_{j}^{m}\rangle
 &=& g_{n}\langle n_{i}^{m}n_{j}^{m}\rangle_{0},
\end{eqnarray}
where $\langle\cdots\rangle$ and $\langle\cdots\rangle_{0}$
represent the expectation values in terms of
$P_{G}|\Phi\rangle$ and $|\Phi\rangle$, respectively.
The renormalization factors $g_{t}, ~g_{s}$ \cite{Zhang2} and $g_{n}$
\cite{Yokoyama} are 
determined by the ratios of the probabilities of the corresponding
physical processes in the states $P_{G}|\Phi \rangle$ and $|\Phi \rangle$:
\begin{equation}
g_{t} = \frac{2\delta}{1+\delta},
~g_{s} = \frac{4}{(1+\delta )^{2}},
~g_{n} = 1,
\end{equation}
where $\delta = 1 - n$ is the doping rate.
Since the double occupancy is excluded in each layer, we can naturally
assume that the interlayer hopping term is also estimated as
\begin{equation}
\langle c^{m\pm 1\dag}_{i\sigma}c_{j\sigma}^{m} \rangle
= g_{t}\langle c^{m\pm 1\dag}_{i\sigma}c_{j\sigma}^{m}\rangle_{0}.
\end{equation}

Next, we consider the relation among the superconducting phases of the layers.
We introduce order parameters in the $m$-th layer as
\begin{equation}
 \Delta_{\tau}^{m} = \langle c_{i+\tau,\uparrow}^{m\dag}
 c_{i,\downarrow}^{m\dag}\rangle_{0},
\end{equation}
with $\tau = x$ and $y$, $i+\tau$ denotes the nearest neighbor of $i$ in 
the $\tau$ direction.
Since we consider the $d_{x^{2}-y^{2}}$-wave symmetry,
$\Delta_{x}^{m} = -\Delta_{y}^{m}$.
When the Josephson current flows along the $z$-direction, the order parameters
satisfy the following phase relation:
\begin{equation}
\Delta_{\tau}^{m} = \Delta_{\tau}^{0}\exp (im\phi).
\end{equation}
Thus, the fermion operator is subject to the following condition:
\begin{equation}
 c_{i\sigma}^{m\pm 1\dag} = \exp(\pm\frac{i}{2}\phi)c_{i\sigma}^{m\dag}.
\end{equation}

Using the Gutzwiller approximation and eq.(9), we can obtain the effective
mean-field Hamiltonian of (3) in momentum space as
\begin{eqnarray}
{\cal H}_{\rm eff} &=& -\sum_{k,\sigma}
T_{k}c_{k\sigma}^{\dag}c_{k\sigma} + N_{s}\epsilon_{0} \nonumber \\
& & - \sum_{k}\left[ \Delta_{\ktd}c_{-k\downarrow}c_{k\uparrow}
 + \Delta_{\ktd}^{\ast}
c_{k\uparrow}^{\dag}c_{-k\downarrow}^{\dag} \right], 
\end{eqnarray}
where $k = (k_{x}, k_{y}, k_{z})$ and 
$\ktd = (k_{x}, k_{y})$ 
are the three- and 
two-dimensional wavevectors, respectively, 
$\ktd = (k_{x}, k_{y})$ is the two-dimensional wavevectors,
$N_{s}$ is the total number of sites, 
\begin{eqnarray*}
T_{k} &=& (g_{t}t+J_{1}\xi)\gamma_{\ktd} 
 - 2g_{t}t_{z}\cos(k_{z}a_{z}-\frac{\phi}{2})
\nonumber \\
& & + J_{N}n + \mu, \nonumber \\
\Delta_{\ktd} &=& J_{2}|\Delta_{\tau}^{0}|\eta_{\ktd}, \nonumber \\
\epsilon_{0} &=& 4J_{2}|\Delta_{\tau}^{0}|^{2} + 4J_{1}\xi^{2}
+ \frac{J_{N}}{2}n^{2}, \nonumber \\
\gamma_{\ktd} &=& 2(\cos(k_{x}a_{x}) + \cos(k_{y}a_{y}) ), \nonumber \\
\eta_{\ktd} &=& 2(\cos(k_{x}a_{x}) - \cos(k_{y}a_{y}) ), \nonumber \\
J_{1} &=& \frac{3}{4}g_{s}J - \frac{1}{4}J_{N},
~~~J_{2} = \frac{3}{4}g_{s}J + \frac{1}{4}J_{N}, \nonumber 
\end{eqnarray*}
and $a_{x}, a_{y}, a_{z}$ are the lattice constants.

This Hamiltonian can be diagonalized using a standard Bogoliubov transformation:\begin{eqnarray}
{\cal H}_{\rm eff} &=& \sum_{k}\left( E_{k} - 2g_{t}t_{z}\sin k_{z}a_{z}\sin
\frac{\phi}{2}\right)\left( A_{k}^{\dag}A_{k} + B_{-k}^{\dag}B_{-k} \right)
\nonumber \\
& & + \sum_{k}( \xi_{k} - E_{k} ) + N_{s}\epsilon_{0},
\end{eqnarray}
where
\begin{equation}
E_{k} = \sqrt{\xi_{k}^{2} + |\Delta_{\ktd}|^{2} }, 
~~~\xi_{k} = -\frac{1}{2}( T_{k} + T_{-k} ) 
\end{equation}

The parameters $\Delta_{\tau}^{0}, ~\xi, ~\mu$ are determined by
the following self-consistent equations:
\begin{eqnarray}
1 &=& \frac{J_{2}}{4N_{s}}\sum_{k}\frac{\eta_{\ktd}^{2}}
{2E_{k}},
\\
\xi &=& -\frac{1}{4N_{s}}\sum_{k}\frac{\xi_{k}\gamma_{\ktd}}
{2E_{k}},
\\
\delta &=&  \frac{1}{N_{s}}\sum_{k}\frac{\xi_{k}}{E_{k}}.
\end{eqnarray}
Then we obtain the ground state energy per site as the function of $\phi$;
\begin{eqnarray}
E_{0}(\phi)/N_{s} &=& \frac{1}{N_{s}}\sum_{k}(\xi_{k} - E_{k}) \nonumber \\
& & + 4J_{2}|\Delta_{\tau}^{0}|^{2} + 4J_{1}\xi^{2} + \frac{J_{N}}{2}n^{2}.
\end{eqnarray} 
In the following calculation, we take $J/t = J_{N}/t = 0.2$  and 
~$t_{z}/t = 0.1$.
%

First, we look at the $\phi$-dependence of $E_{0}(\phi)/N_{s}$.
In order to see the variation in the energy clearly, we define
$\idel E(\phi) = ( E_{0}(\phi) - E_{0}(0) )/N_{s}$. 
In Fig.2, $\it{\Delta} E(\phi)$ is plotted against the phase $\phi$ for 
various values of the doping rate $\delta$.
We can see that $\it{\Delta} E(\phi)$ is the monotonic increasing
 function of $\phi$,
and that $\it{\Delta} E(\phi)$ increases with $\delta$ for arbitrary
 values of $\phi$.
When $\delta\geq 0.25$ ( the line (e) ), however, $E(\phi)$ exceeds
the energy of the normal state, i.e., the superconducting state becomes unstable
in the region $\phi\geq 0.3\pi$.
Let us discuss this situation a little more.
From eq.(6), we can interpret $g_{t}t_{z}$ as the effective amplitude
of the interlayer hopping.
Thus, as $\delta$ increases, the coupling strength between layers increases.
On the other hand, the gap amplitude $\Delta_{\tau}^{0}$ 
linearly decreases with
increasing of $\delta$ \cite{Zhang2}.
If the gap amplitude $\Delta_{\tau}^{0}$ 
is smaller than the effective interlayer
hopping amplitude $g_{t}t_{z}$, the quasiparticle energy
\begin{equation}
 \sqrt{\xi_{k}^{2}+|\Delta_{\ktd}|^{2}}
 - 2g_{t}t_{z}\sin k_{z}a_{z}\sin\frac{\phi}{2}
\end{equation}
becomes
negative for $\phi\geq\phi_{c}$, where $\phi_{c}$ is a critical phase.
Consequently, the superconducting state is no more stable 
as compared to the normal state. 
\begin{figure}
\vspace{20pt}
\begin{center}
\leavevmode
 \epsfxsize=80mm
 \epsfbox{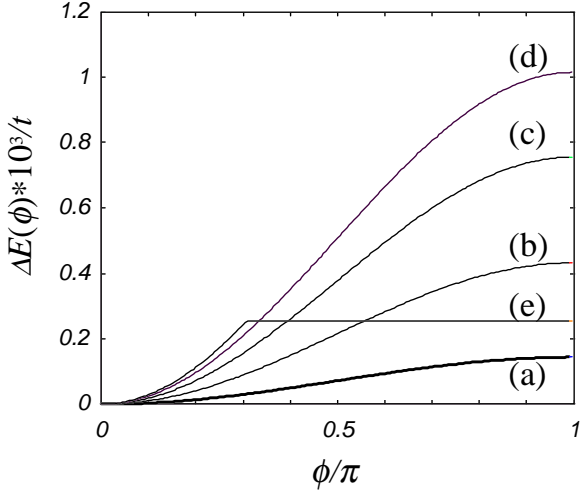}
\end{center}
\caption{
The $\phi$ dependence of $\idel(\phi)$ 
for (1) $t_{z}/t=0.1$. 
We take (a) $\delta=0.05$, (b) $\delta=0.1$, (c) $\delta=0.15$, 
(d) $\delta=0.2$ and (e) $\delta=0.25$.
}
\end{figure}

Next, we calculate the Josephson current defined as
\begin{equation}
j(\phi) = \frac{2e}{\hbar}\frac{\partial}{\partial\phi}E_{0}(\phi).
\end{equation}
Figure 2 shows the Josephson current-phase relations 
obtained for
various values of $\delta$.
The currents show the sinusoidal $\phi$-dependence
as expected 
and have their maximum values $j_{\rm max}(\delta)$ at
$\phi = \frac{\pi}{2}$.
When $\delta\leq0.2$, $j_{\rm max}$ increases with $\delta$.
This is because the effective mass of a Cooper pair, that is,
its localization tendency in a layer decreases in rough proportion to
$1/|\Delta_{\tau}^{0}|$ \cite{Tanaka},
with $|\Delta_{\tau}^{0}|$ being the decreasing function of $\delta$.
On the contrary, when $\delta\geq0.25$, $j_{\rm max}$
decreases with increasing of $\delta$ because the Josephson current
vanishes for $\phi\geq\phi_{c}$.
Thus $j_{\rm max}$ reaches its maximum in the region
$0.2\leq\delta\leq0.25$.
In order to see this situation in more detail,
the doping dependence of $j_{\rm max}$ is plotted explicitly in Fig.3.
\begin{figure}
\vspace{20pt}
\begin{center}
\leavevmode
 \epsfxsize=80mm
 \epsfbox{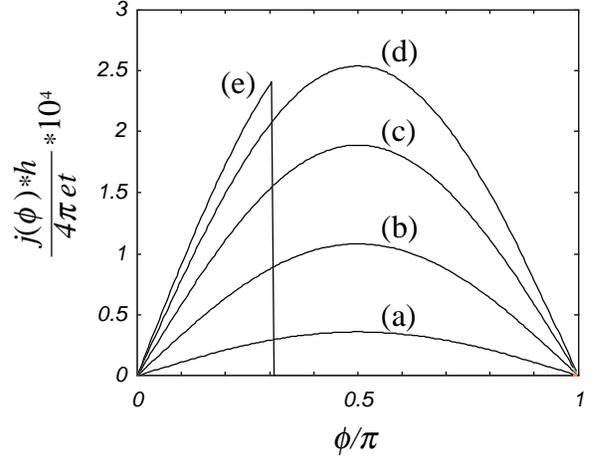}
\end{center}
\caption{
The $\phi$ dependence of the Josephson current $j(\phi)/t$ 
for $t_{z}/t=0.1$. 
We take (a) $\delta=0.05$, (b) $\delta=0.1$, (c) $\delta=0.15$, 
(d) $\delta=0.2$ and (e) $\delta=0.25$.
}
\end{figure}
\begin{figure}
\vspace{20pt}
\begin{center}
\leavevmode
 \epsfxsize=80mm
 \epsfbox{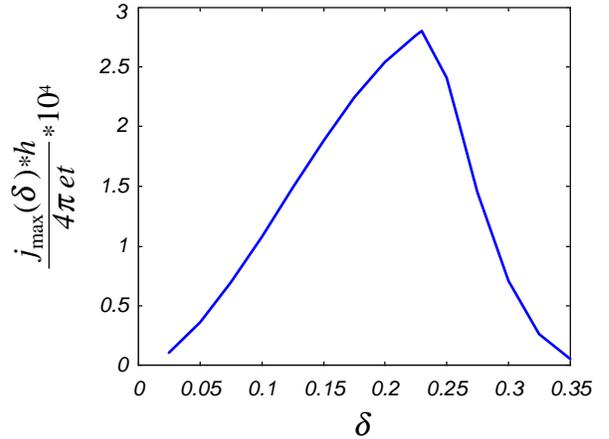}
\end{center}
\caption{
The maximum Josephson current $j_{\rm max}(\delta)/t$ plotted as
a function of $\delta$ for $t_{z}/t=0.1$.
}
\end{figure}
We can see that $j_{\rm max}$ has its maximum value
at $\delta\simeq 0.23$ in the present case. \par
In the recent experiment\cite{Matsuda2}, 
it has been reported that the intrinsic
Josephson current increases with hole-doping (in our notation, $\delta$)
when the system is in the under-doped region.
We note here that, if we do not assume 
the doping dependence of interlayer hopping 
as in Eq. (10), $i.e.$, 
the interlayer hopping amplitude has no doping dependence, 
$j(\phi)$ becomes a decreasing function of $\delta$.
Since the renormalization of interlayer hopping  in 
eq.(10) is a natural assumption to treat the layered $t$-$J$ model 
based on the Gutzwiller approximation, 
the doping dependence of the intrinsic Josephson current observed 
in actual experiments can be used 
to check the validity of the Gutzwiller approximations 
of the layered $t$-$J$ model.
Of course there is only a few experiments concerning a doping dependence of the
Josephson current so far, so that more extensive
comparison between the theory and experiments is necessary.

In this letter, we have calculated the d.c. Josephson 
 current which flows perpendicular to the 
CuO$_{2}$ planes in high $T_{C}$ superconductors using the 
layered $t$-$J$ model within the mean field theory and Gutzwiller 
approximation. The maximum Josephson current has its maximum 
near the optimum doping rate.
This is due to the competition between the enhancement of the 
effective transfer energy along the $c$-axis and the 
decrease of the magnitude of the order parameter 
with the increase of the doping rate. 
We hope such a behavior will be observed in experiments near future.

\section*{Acknowledgment}
The authors wish to thank Y. Matsuda and M. Ogata for useful discussions.
This work is supported by
the Core Research for
Evolutional Science and technology (CREST) of the Japan Science
and Technology Corporation (JST). 
One of the author Y.T. is supported by 
Grant-in Aid for Scientific Research from the 
Ministry of Education, Science, Sports and Culture. 
The computational aspect of this work has been performed at the 
facilities of the Yukawa Institute. 

\end{document}